\documentclass[
  aip,
  jap,
  twoside,
  twocolumn,
  10pt,
  floatfix,
  showpacs,
  preprintnumbers,
]{revtex4-1}

\usepackage[final]{graphicx}
\usepackage{tabularx}
\usepackage[nooneline]{subfigure}
\usepackage{amsmath,amssymb}
\usepackage{dcolumn}
\usepackage{color}

\newcommand{\etal}[0]{\textit{et al.}}

\newcommand{\VB}[0]{\text{VB}}
\newcommand{\VBM}[0]{\text{VBM}}
\newcommand{\CB}[0]{\text{CB}}

\newcommand{\expt}[0]{\text{expt}}
\newcommand{\calc}[0]{\text{calc}}
\newcommand{\BTO}[0]{\text{BaTiO$_3$}}
\newcommand{\Ba}[0]{\text{Ba}}
\newcommand{\Ti}[0]{\text{Ti}}
\renewcommand{\O}[0]{\text{O}}
\newcommand{\VBa}[0]{\text{{\it V}$_{\Ba}$}}
\newcommand{\VTi}[0]{\text{{\it V}$_{\Ti}$}}
\newcommand{\VO}[0]{\text{{\it V}$_{\O}$}}
\newcommand{\VBaVO}[0]{\text{\VBa--\VO}}
\newcommand{\VTiVO}[0]{\text{\VTi--\VO}}
\newcommand{\K}[0]{\text{K}}
\newcommand{\eV}[0]{\text{eV}}

\newcommand{\nm}[0]{\text{nm}}

\newcommand{\atm}[0]{\text{atm}}

\newcommand{\sect}[1]{Sect.~\ref{#1}}
\newcommand{\fig}[1]{Fig.~\ref{#1}}
\newcommand{\eq}[1]{(\ref{#1})}
\newcommand{\tab}[1]{Table~\ref{#1}}
\newcommand{\tbhd}[1]{\textbf{#1}}

\renewcommand{\epsilon}[0]{\varepsilon}

\begin{document}

\title{
  Thermodynamics of mono and di-vacancies in barium titanate
}

\date{\today}

\author{Paul Erhart}
\email{paul.erhart@web.de}%
\author{Karsten Albe}
\affiliation{
  Institut f\"ur Materialwissenschaft,
  Technische Universit\"at Darmstadt,\\
  Petersenstra{\ss}e~23,
  D-64287 Darmstadt,
  Germany
}

\preprint{published in Journal of Applied Physics {\bf 102}, 084111 (2007), doi:10.1063/1.2801011}

\begin{abstract}
The thermodynamic and kinetic properties of mono and di-vacancy
defects in cubic (para-electric) barium titanate are studied by means of
density-functional theory calculations. It is determined which vacancy
types prevail for given thermodynamic boundary conditions. The
calculations confirm the established picture that vacancies occur in
their nominal charge states almost over the entire band gap. For the
dominating range of the band gap the di-vacancy binding energies are
constant and negative. The system, therefore, strives to achieve a
state in which under metal-rich (oxygen-rich) conditions all metal
(oxygen) vacancies are bound in di-vacancy clusters. The migration barriers
are calculated for mono-vacancies in different charge states. Since oxygen
vacancies are found to readily migrate at typical growth temperatures,
di-vacancies can be formed at ease. The key results of the present study with
respect to the thermodynamic behavior of mono and di-vacancies influence the
initial defect distribution in the ferroelectric phases and therefore the
conditions for aging.
\end{abstract}


\maketitle

\section{Introduction}

Barium titanate is a prototypical ferroelectric material.
At ambient pressure and temperatures above 393\,K it assumes a para-electric,
cubic perovskite structure. At lower temperatures a tetragonal distortion of
the unit cell is observed which gives rise to ferro-electricity. At even lower
temperatures further symmetry reductions lead to ferroelectric orthorhombic
and rhombohedral phases. The sequence of phase transitions is the result of
subtle structural and energetic differences. \cite{Coh92, ZhoVanKin95,
  GhoGonMic97, TinStaSep00} Probably the most important technological
application of barium titanate is in thin-film
capacitors. \cite{Smy00} It also serves as an end member in several
lead-free ferroelectric alloys, \cite{FukLiUes03} and is used in
combination with SrTiO$_3$ to obtain tunable RF
devices. \cite{Tombak2002} Because of its technological importance and
its standing as a prototypical ferroelectric, barium titanate has been
intensively investigated both experimentally and theoretically.

Intrinsic and extrinsic point defects are of special importance in
semi-conductors as they control to a great extent the electronic
properties of these materials. In case of barium titanate, this has
motivated a considerable number of studies which used conductivity
\cite{DanHar76, EroSmy78, ChaShaSmy81, ChaShaSmy82, ChaSmy84,
  SonYoo99, Smy00} and diffusivity \cite{SonYoo99, SonYoo00}
measurements to infer information about the point defect equilibria in
this material. It is furthermore assumed that oxygen vacancies play a key
role in ageing and fatigue of ferroelectric materials  by impeding
domain wall motion or by acting as local disturbances of the
polarization (see e.g., Refs.~\onlinecite{NeuArl87, ArlNeu88,
  WarDimTut95, HeVan03, LiMaChe05}). For this reason, oxygen vacancies
and their associates have been intensively studied in experimentally
(see e.g., Refs.~\onlinecite{WarVanDim96, KeeNieKri98, ZafJonJia98})
as well as theoretically (see e.g., Refs.~\onlinecite{ParCha98,
  PoyCha99a, PoyCha99b, PoyCha00b, PoyCha00a, Par03, DuqSta03,
  CocBur04, ShiWad04, ShiWad05}).

Experimental investigation of point defect properties, however, are 
usually restricted in that they either provide bulk
information (e.g., conductivity, diffusivity), which corresponds to a
macroscopic average over the sample and is therefore defect unspecific, or
very localized information (e.g., electron spin resonance, positron
annihilation spectroscopy), which is, however, restricted to certain charge
states (unpaired spin states) or types of defects (open volumes,
vacancies). \cite{AllLid03} In general, the correlation between
experimental data and specific defects is indirect and therefore
potentially ambiguous. Quantum mechanical calculations on the other hand are
capable of providing a very detailed picture of individual defects and
simultaneously sample a whole variety of both different configurations
and charge states. They are therefore a very powerful tool for
verifying model assumptions and for providing fundamental insights
into the energetics of defects as well as their kinetic and electronic
properties.

The purpose of the present work is twofold. First, we aim to verify
the defect models which have been employed to explain the experimental
observations (in particular conductivity and diffusivity
measurements). Second, we explore the formation of di-vacancies which
-- as argued above -- is an important ingredient for understanding the
deterioration of ferroelectric switchability.
%NEW: begin
The calculations are carried out for the cubic phase which is the most
stable phase at high temperatures at which most ceramics are processed
and at which the initial defect distributions are installed.
%NEW: end

\section{Methodology}

\subsection{Computational setup}

Calculations were carried out within density--func\-tional theory
(DFT) using the Vienna ab-initio simulation package (VASP)
\cite{KreHaf93, KreHaf94, KreFur96a, KreFur96b} in combination with
the pro\-jec\-tor-aug\-mented wave (PAW) method to represent the ionic
cores and core electrons. \cite{Blo94, KreJou99} In order to find the
most suitable representation for the exchange-correlation (XC)
functional preliminary calculations were performed within the
local-density approximation (LDA) \cite{CepAld80, PerZun81} as well as
the generalized-gradient approximation (GGA) in the
Perdew-Burke-Ernzerhof (PBE) parameterization. \cite{PerBurErn96} We
furthermore considered the effect of treating the Ti-3s and Ti-3p
electrons as semi--core states. All calculations were carried out with
20 $k$-points for Brillouin zone sampling and a plane wave--
cutoff energy of 500\,eV. For each setup the energy-volume curve of
cubic barium titanate was computed and the equilibrium lattice
constant, cohesive energy, bulk modulus and its pressure derivative
were subsequently obtained from a fit to the Birch-Murnaghan equation
of state. Finally, the full band diagrams were calculated at the
respective zero pressure lattice constant.

\begingroup
%\squeezetable
\begin{table*}
  \caption{
    Bulk properties of cubic barium titanate as obtained from
    experiment and first-principles calculations.
    US-PP: ultrasoft pseudo-potentials;
    FP-LAPW: full potential-linearized augmented plane waves;
    TB-LMTO: tight-binding linear muffin-tin orbitals;
    ASA: atomic sphere approximation;
    LDA: local-density approximation;
    GGA: generalized-gradient approximation;
    PBE: Perdew-Burke-Ernzerhof parameterization of the GGA;
    $E_c$: cohesive energy (eV/f.u.);
    $a_0$: lattice constant (\AA);
    $V_0$: equilibrium volume (\AA$^3$/f.u.);
    $B$, $B'$: bulk modulus (GPa) and its pressure derivative;
    $E_G^{\Gamma-\Gamma}$: direct band gap at $\Gamma$-point (eV);
    $E_G^{\text{R}-\Gamma}$: indirect band gap measured between points
    R and $\Gamma$;
    $m^*_e$, $m^*_h$: effective electron (hole) mass at the
    $\Gamma$-point along $\left<100\right>$ in units of the electron
    mass.
  }
  \label{tab:methods_comparison}
  \centering

  \newcommand{\spacer}[0]{\text{\phantom{\hspace{0.3cm}}}}
  \newcolumntype{C}{>{\centering\centering\arraybackslash}X}

  \begin{minipage}{\linewidth}
    \begin{tabularx}{\linewidth}{lcccccl*{6}{C}}
      \hline\hline
      & \tbhd{Expt.}
      & \multicolumn{4}{c}{\tbhd{DFT (Literature)}}
      & \phantom{\hspace{0.1cm}}
      & \multicolumn{6}{c}{\tbhd{DFT (This work)}} \\[3pt]
      \cline{3-6}
      \cline{8-13}
      &
      & Ref.~\onlinecite{UluCagGod02}
      & Ref.~\onlinecite{SalHosSha04}
      & Ref.~\onlinecite{SalHosSha04}
      & Ref.~\onlinecite{SahSinMoo00}
      &
      & \multicolumn{3}{c}{GGA-PBE}
      & \multicolumn{3}{c}{LDA} \\
      &
      & US-PP & FP-LAPW & FP-LAPW & TB-LMTO
      &
      & Ti      & Ti-3p      & Ti-3s-3p
      & Ti      & Ti-3p      & Ti-3s-3p \\
      &
      & GGA & LDA & GGA & ASA, LDA \\

      \hline

      $E_c$
      & $-31.57$ \footnote{as cited in Reference~\onlinecite{UluCagGod02}}
      & $-37.92$ & & & &
      & $-39.62$ & $-39.98$ & $-40.06$ & $-43.84$ & $-44.10$ & $-44.11$ \\

      $a_0$
      & 3.992
      & 4.006 & 3.9   & 4.0   &
      &
      & 4.038 & 4.037 & 4.038 & 3.957 & 3.953 & 3.955 \\

      $V_0$
      &
      & 64.28 & 61.0  & 62.5 -- 65.2  &
      &
      & 65.84 & 65.79 & 65.84 & 61.98 & 61.79 & 61.86 \\

      $B$
      & 173\footnote{as cited in Reference~\onlinecite{SalHosSha04}}
      & 168 & 146 & 185 -- 189 &
      &
      & 165 & 161 & 160 & 200 & 194 & 193 \\

      $B'$
      &
      & 4.5\footnote{fitted to Rose's equation of state \cite{RosSmiGui84}} & & &
      &
      & 4.7 & 4.5 & 4.5 & 4.7 & 4.6 & 4.6 \\[6pt]

      $E_G^{\Gamma-\Gamma}$
      & 3.27, 3.38\footnote{Reference~\onlinecite{Wem70},
	obtained from optical measurements on tetragonal barium
	titanate at room temperature.}
      & & 1.8 & 1.85 -- 1.9 & 1.2\footnote{at the experimental lattice
	constant}
      &
      & 1.69 & 1.81 & 1.85 & 1.68 & 1.80 & 1.82 \\

      $E_G^{\text{R}-\Gamma}$
      &
      & & & & &
      & 1.56 & 1.66 & 1.69 & 1.62 & 1.71 & 1.73 \\[6pt]

      $m_e^*$
      &
      & & & & &
      & 1.16 & 1.16 & 1.16 & 1.01 & 1.00 & 1.01 \\
      & & & & & &
      & 16.7 & 15.8 & 15.6 & 13.5 & 12.8 & 12.7 \\

      $m_h^*$
      &
      & & & & &
      & 0.96 & 0.97 & 0.97 & 0.89 & 0.89 & 0.89 \\
      & & & & & &
      & 3.24 & 3.31 & 3.32 & 2.88 & 2.93 & 2.94 \\

      \hline\hline
    \end{tabularx}
  \end{minipage}
\end{table*}
\endgroup

The results of these preparatory calculations are compiled in
\tab{tab:methods_comparison} in comparison with experimental and
theoretical data from literature. The influence of the Ti-3p and
Ti-3s states on the properties included in this comparison is very
small. The most significant difference is the increase of the cohesive
energies by about 1\% upon inclusion of the Ti-3s and Ti-3p electrons
in the valence. Thus, while for some properties the deep Ti-3s and
Ti-3p can play a crucial role, in the present context their effect is
expected to be small. For the sake of computational efficiency
we did, therefore, not include the Ti-3s and Ti-3p electrons in the
valence.

The exchange-correlation functional on the other hand has a more
pronounced impact. Within the GGA the bulk modulus is reasonably well
reproduced, but the lattice constant is overestimated while the
opposite applies for the LDA results. These findings are consistent
with the results of previous studies. \cite{GhoGonMic99, UluCagGod02,
  WuCohSin04} In the present study we have decided to employ the LDA.

\begingroup
%\squeezetable
\begin{table}
  \caption{
    Bulk properties of Ba, Ti and O and their compounds in their
    respective ground-states. Experimental data from
    References~\onlinecite{Kit04, crchandbook, EveMcC92}.
    $E_c$: cohesive energy (eV/atom);
    $c/a$: axial ratio;
    $r_0$: dimer bond length (\AA);
    $\Delta H_f$: enthalpy of formation (eV/f.u.);
    other symbols as in \tab{tab:methods_comparison}.
  }
  \label{tab:phases}
  \centering

  \newcommand{\struct}[1]{\multicolumn{3}{l}{\textbf{#1}}}
  \newcommand{\myhline}{\hline\vspace{-8pt}\\}
  \newcommand{\mycline}[1]{\cline{#1}\vspace{-8pt}\\}
  \newcommand{\spacer}[0]{\text{\phantom{\hspace{0.3cm}}}}

  \newcolumntype{C}{>{\centering\centering\arraybackslash}X}

  \begin{tabular}{lcc}
    \hline\hline
    & \multicolumn{1}{c}{\tbhd{Experiment}}
    & \multicolumn{1}{c}{\tbhd{This work}} \\[3pt]
    
    \myhline
    \struct{Ba, body-centered cubic
      (Im$\mathbf{\bar{3}}$m, no. 229, A2)} \\[3pt]
    \spacer $E_c$ & $-1.87$, $-1.90$ & $-1.12$ \\
    \spacer $a_0$ & 5.020 & 4.770 \\
    \spacer $B$   & 10 & 5.3 \\[3pt]
    
    \struct{Ti, hexagonal-close packed
      (P$\mathbf{6_3}$/mmc, no. 194, A3)} \\
    \spacer $E_c$ & $-4.85$ & $-8.53$ \\
    \spacer $a_0$ & 2.950 & 2.852 \\
    \spacer $c/a$ & 1.588 & 1.585 \\
    \spacer $B$   & 110   & 133 \\[3pt]
    
    \struct{O, dimer} \\[3pt]
    \spacer $E_c$ & $-2.583$ & $-4.79$ \\
    \spacer $r_0$ & 1.21 & 1.22 \\[3pt]
    
    \struct{BaO, cesium chloride
      (Pm$\mathbf{\bar{3}}$m, no. 221, B2)} \\[3pt]
    \spacer $\Delta H_f$ & $-5.68$ & $-6.68$ \\
    \spacer $a_0$ & 5.496 & 3.281 \\
    \spacer $B$   & & 96 \\[3pt]
    
    \struct{BaO$_{\mathbf{2}}$
      (I4/mmm, no. 139, C11$_{\mathbf{b}}$)} \\
    \spacer $\Delta H_f$ & & $+0.95$ \\
    \spacer $a_0$ & 3.78 -- 3.81 & 3.768 \\
    \spacer $c/a$ & 1.79 & 1.783 \\
    \spacer $B$   & & 44 \\[3pt]
    
    \struct{TiO$_{\mathbf{2}}$, rutile
      (P4$_{\mathbf{2}}$/mnm, no. 139, C4)} \\
    \spacer $\Delta H_f$ & $-9.78$ & $-10.93$ \\
    \spacer $a_0$ & 4.594 & 4.572 \\
    \spacer $c/a$ & 0.644 & 0.640 \\
    \spacer $B$   & 210 & 242 \\[3pt]
    
    \struct{BaTiO$_{\mathbf{3}}$, cubic perovskite
      (Pm$\bar{\mathbf{3}}$m, no. 221, E2$_{\mathbf{1}}$)} \\
    \spacer $\Delta H_f$ & $-20.84$ & $-19.83$ \\
    \spacer $a_0$ & 3.996 & 3.957 \\
    \spacer $B$   & 162, 167 & 200 \\
    \hline\hline
  \end{tabular}
\end{table}
\endgroup

In order to determine the properties of the phases in
\tab{tab:phases} we have employed the same computational settings as
for BaTiO$_3$ but varied the number of $k$-points for each material
such as to ensure a convergence of the total energy better than
1\,meV/unit cell. Energy-volume curves were calculated for each
of these structures allowing for full internal relaxation.

\subsection{Defect calculations}

\subsubsection{Formation energies}
\label{sect:eform_calc}

The formation energy $\Delta E_D^f$ of a defect in charge state $q$ is
given by \cite{QiaMarCha88, ZhaWeiZun98a, ZhaWeiZun98b, ZhaPerLan04}
\begin{align}
  \Delta E_D^f
 &= (E_D - E_H) + q (E_{\VBM} + E_F) - \sum_i \Delta n_i \mu_i
  \label{eq:eform}
\end{align}
where $E_D$ is the total energy of the defective system and $E_H$ is
the total energy of the perfect reference cell. The second term
describes the dependence on the Fermi level, $E_F$, where $E_{\VBM}$
is the position of the valence band maximum. The variation of the
formation energy with the chemical potentials of the constituents is
given by the last term. The difference between the number of atoms of
type $i$ in the reference cell with respect to the defective cell is
denoted by $\Delta n_i$. The chemical potential $\mu_i$ 
of constituent $i$ can be rewritten as
 $\mu_i = \mu_i^{bulk} + \Delta\mu_i$ where $\mu_i^{bulk}$
denotes the chemical potential of the standard reference
state and is equivalent to the cohesive energy per atom(see
\sect{sect:phasediagram}).

\begin{figure}
  \centering
  \includegraphics[width=0.8\columnwidth]{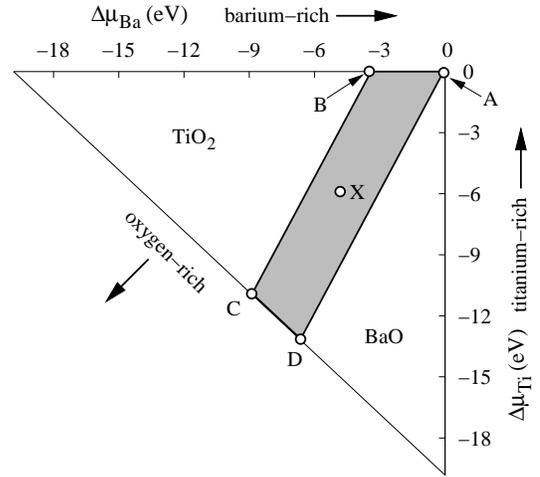}
  \caption{
    Stability diagram for cubic barium titanate as determined from
    density-functional theory calculations. The area confined between
    points A, B, C and D is the chemical stability range of \BTO.
    The line through points C and D corresponds to maximally oxygen-rich
    conditions and an oxygen chemical potential of
    $\Delta\mu_{\O}=0\,\eV$. Along lines parallel to C--D the oxygen
    chemical potential is constant. The most negative value of
    $\Delta\mu_{\O}=\Delta H_f/3=-6.61\,\eV$ is obtained in the upper
    right corner of the diagram.
  }
  \label{fig:phasediagram}
\end{figure}

\begingroup
%\squeezetable
\begin{table}
  \caption{
    Formation energies of mono and di-vacancies under the chemical
    conditions indicated in \fig{fig:phasediagram}.
    Note that if $\Delta\mu_{\Ba}$ and $\Delta\mu_{\Ti}$ are given
    $\Delta\mu_{\O}$ is uniquely determined by equation
    \eq{eq:Hf_BTO}.
    %NEW: begin
    The charge state, $q$, of the defect which determines the Fermi
    level dependence of the formation energies via equation
    \eq{eq:eform} is given in the second column. 
    The number of electrons occupying conduction band states ($n_e$)
    and holes occupying valence band states ($n_h$) are relevant for the
    band gap correction via equation \eq{eq:bandgap_correction}
    and are given in the third column where positive and negative
    values indicate $n_h$ and $-n_e$, respectively.
    %NEW: end
    All energies are given in units of eV.
    The finite-size scaling extrapolation error is given in the last
    column.
  }
  \label{tab:eform}
  \centering

  \newcommand{\spread}[1]{\multicolumn{1}{l}{#1}}

  \newcolumntype{C}{>{\centering\centering\arraybackslash}X}
  \newcolumntype{e}[0]{>{\centering\centering\arraybackslash}D{.}{.}{4.2}}
  \newcolumntype{f}[1]{>{\centering\centering\arraybackslash}D{.}{.}{#1.2}}
  \newcolumntype{g}[0]{>{\centering\centering\arraybackslash}D{.}{.}{2.0}}
  \newcolumntype{Q}[0]{>{(}d<{)}}

  \begin{tabular}{lgge*{3}{e}Q}
    \hline\hline
    \multicolumn{1}{l}{\tbhd{Defect}}
    & \multicolumn{1}{c}{\tbhd{$\boldsymbol{q}$}}
    & \multicolumn{1}{c}{\tbhd{$\boldsymbol{n_{e,h}}$}}
    & \multicolumn{1}{c}{\tbhd{X}}
    & \multicolumn{1}{c}{\tbhd{A}}
    & \multicolumn{1}{c}{\tbhd{C}}
    & \multicolumn{1}{c}{\tbhd{D}}
    & \multicolumn{1}{c}{\tbhd{Err.}}\\[3pt]
    
    \hline
    \multicolumn{1}{l}{$\Delta\mu_{\Ba}$}
    &&
    & \multicolumn{1}{e}{-4.78}
    & \multicolumn{1}{e}{-0.10}
    & \multicolumn{1}{e}{-8.90}
    & \multicolumn{1}{e}{-6.68} \\
    
    \multicolumn{1}{l}{$\Delta\mu_{\Ti}$}
    &&
    & \multicolumn{1}{e}{-6.02}
    & \multicolumn{1}{e}{0.00}
    & \multicolumn{1}{e}{-10.93}
    & \multicolumn{1}{e}{-13.15} \\
    
    \multicolumn{1}{l}{$\Delta\mu_{\O}$}
    &&
    & \multicolumn{1}{e}{-3.01}
    & \multicolumn{1}{e}{-6.58}
    & \multicolumn{1}{e}{0.00}
    & \multicolumn{1}{e}{0.00} \\[3pt]
    
    \hline
    
    \spread{$\VO$}
    & 0  &  -2 &  5.21 &  1.64 &   8.22 &   8.22 &   0.07 \\
    & +1 &  -1 &  2.08 & -1.49 &   5.09 &   5.09 &   0.14 \\
    & +2 &   0 & -1.50 & -5.07 &   1.51 &   1.51 &   0.08 \\[6pt]
    
    \spread{$\VBa$}
    & -2 &  +2 &  5.94 & 10.61 &   1.82 &   4.04 &   0.05 \\
    & -1 &  +1 &  5.68 & 10.36 &   1.56 &   3.78 &   0.03 \\
    & 0  &   0 &  5.57 & 10.25 &   1.45 &   3.67 &   0.02 \\[6pt]
    
    \spread{$\VBaVO$}
    & -2 &  -2 & 10.97 & 12.08 &   9.86 &  12.08 &   0.14 \\
    & -1 &  -1 &  7.26 &  8.37 &   6.15 &   8.37 &   0.14 \\
    & 0  &   0 &  3.83 &  4.94 &   2.72 &   4.94 &   0.13 \\
    & +1 &  +1 &  3.77 &  4.88 &   2.66 &   4.88 &   0.11 \\[6pt]
    
    \spread{$\VTi$}
    & -4 &   0 &  9.33 & 15.35 &   4.42 &   2.19 &   0.15 \\
    & -3 &  +1 &  8.92 & 14.94 &   4.01 &   1.79 &   0.10 \\
    & -2 &  +2 &  8.69 & 14.71 &   3.78 &   1.56 &   0.11 \\
    & -1 &  +3 &  8.56 & 14.58 &   3.65 &   1.42 &   0.08 \\
    & 0  &  +4 &  8.53 & 14.55 &   3.62 &   1.40 &   0.07 \\[6pt]
    
    \spread{$\VTiVO$}
    & -4 &  -2 & 13.24 & 15.70 &  11.35 &   9.12 &   0.22 \\
    & -3 &  -1 &  9.46 & 11.91 &   7.56 &   5.34 &   0.22 \\
    & -2 &   0 &  5.91 &  8.37 &   4.01 &   1.79 &   0.21 \\
    & -1 &  +1 &  5.70 &  8.16 &   3.80 &   1.58 &   0.13 \\
    & 0  &  +2 &  5.64 &  8.10 &   3.74 &   1.52 &   0.12 \\
    & +1 &  +3 &  5.66 &  8.12 &   3.76 &   1.54 &   0.10 \\
    
    \hline\hline
  \end{tabular}
\end{table}
\endgroup

For the defect calculations we employed supercells with 40, 60, 90 and
135 atoms equivalent to $2\times 2\times 2$ to $3\times 3\times 3$
unit cells. 
The formation energies given in \tab{tab:eform} were obtained by
extrapolating the data calculated for different supercell sizes to infinite
dilution (see below).
We considered both mono-vacancies (\VO, \VBa, \VTi) as
well as nearest-neighbor di-vacancies (\VBaVO, \VTiVO) taking into
account a variety of charge states (see \tab{tab:eform}).
Due to their size, it is reasonable to assume that host metal ion
interstitials have very large formation energies. This assumption is
supported by analytical potential calculations which have found
Frenkel defects to have much higher energies than Schottky
defects. \cite{LewCat86}
%NEW: begin
In order to confirm these results we calculated the formation energies
of all interstitials and anti-sites in the neutral charge state. Under
metal-rich conditions the formation energies for interstitials and
anti-sites are at least $3.8\,\eV$ larger than for the lowest neutral
vacancy. Under oxygen-rich conditions the difference is smaller but
even for the most favorable case (O on Ti anti-site at point D in
\fig{fig:phasediagram}) the energy difference is at least
0.9\,eV. 
In the following we therefore neglect interstitials and anti-sites and
focus exclusively on vacancy defects.
It should be pointed out that both metal and oxygen interstitials can
play important roles in several other oxides with more open lattice
structures (e.g., TiO$_2$, ZnO, SnO$_2$). In the case of perovskite
lattices the interstitial sites are, however, much smaller and
interstitials lead to significant strains which cause large formation
energies.
%NEW: end

The Brillouin zone integrations were carried
out using a non-shifted $2\times 2\times 2$ $\Gamma$-centered mesh
which depending on the symmetry of the defect configuration is
equivalent to 4 to 6 $k$-points in the irreducible wedge of the
Brillouin zone. The plane wave cutoff--energy was set to 500\,eV.

DFT calculations of point defect formation energies in semi--conductors are
subject to certain shortcomings, which in order to obtain physically
meaningful results must be properly taken into account.
\cite{PerZhaLan05, ErhAlbKle06} Although the band gap is typically
underestimated, energy differences within the valence band and
conduction bands, respectively, are usually rather well described. 
A simple correction of the band structure is, therefore, obtained
by rigidly shifting the valence band ($\Delta E_{\VB}$) and the
conduction band ($\Delta E_{\CB}$) with respect to each other.
The correction energy is then given by
\begin{align}
  \Delta E_{bg} &= n_e \Delta E_{\CB} + n_h \Delta E_{\VB}.
  \label{eq:bandgap_correction}
\end{align}
%OLD:begin
%where $n_e$ and $n_h$ denote the number of electrons and holes in the
%conduction and valence bands, respectively.
%OLD: end
%NEW: begin
For a given defect and charge state the number of electrons in the
conduction band $n_e$ was determined by integrating the number of
occupied conduction band states. The number of holes in the valence
band $n_h$ was similarly obtained as the number of empty valence band
states. 
It should be noted that this correction only considers the effect of
the band gap error on the band energy and assumes rigid levels. It
does not take into account electronic relaxations which occur if
self-interaction effects are properly included.
%NEW: end

Since experimentally the band gap of cubic \BTO\ cannot be  determined
for $T\rightarrow 0\,\K$ due to the ferroelectric phase transition, we
estimated its value by extrapolation of the data at higher
temperatures which yields $E_G^{\expt}=3.4\,\eV$ (compare Fig.~10 in
Ref.~\onlinecite{Wem70}). In order to be able to correct for the
underestimation of the band gap, we furthermore assumed the offset of
the calculated band structure to be restricted to the conduction band,
i.e. $\Delta E_{\VB}=0$ and $\Delta E_{\CB}=E_G^{\expt}-E_G^{\calc}$.
  
Due to the use of supercells elastic and electrostatic interactions
between the periodic images of the defects need to be taken into
account. \cite{MakPay95, ErhAlbKle06} Elastic interactions scale
inversely with the volume, $\mathcal{O}(V^{-1})$, and therefore, have
been corrected using finite-size scaling. For charged defects
electrostatic interactions due to the net charge moment are
present. The corresponding correction term can be expressed in the
form of a multipole expansion. \cite{MakPay95} The leading term, which
describes monopole-monopole interactions, scales with $V^{-1/3}$ and
can be calculated explicitly. The next term is due to
monopole-quadrupole interactions. It scales with $V^{-1}$ and can in
principle also be evaluated explicitly. Since it displays the same
scaling behavior as the elastic interactions, it is, however,
conveniently corrected using the same finite-size scaling procedure.
\cite{MakPay95, LenMozNie02, ErhAlbKle06} In addition, the latter
approach has the advantage to avoid ambiguities in the
calculation of the moments of the net charge distribution. It also allows
to assess the reliability of the correction by means of the error of
the linear extrapolation of the data. For strongly delocalized excess
charge distributions higher order terms might become important which
has, however, not been observed in the present calculations.

Since the magnitude of the el\-ec\-tro\-sta\-tic in\-ter\-ac\-tions in
con\-den\-sed mat\-ter is screened, the mono\-pole-mono\-pole
corr\-ec\-tion term requires know\-ledge of the static dielectric
constant. For reasons of consistency the latter should be taken from
first-principles calculations as well. Using a similar approach as in
the present work Ghosez \etal\ obtained an average value of
$\epsilon=57$ for barium ti\-ta\-nate \cite{GhoGonMic97, Coc03} which
was used in the present work.

\subsubsection{Transition energies}

If the formation energies of a given defect in charge states $q_1$ and
$q_2$ are known, the thermal (equilibrium) transition level can be
obtained according to
\begin{align}
  \epsilon &= -\frac{\Delta E^f_D(q_1)-\Delta E^f_D(q_2)}{q_1-q_2}
  \label{eq:trans}
\end{align}
where $\Delta E^f_D(q_1)$ and $\Delta E^f_D(q_2)$ denote the formation
energies at the valence band maximum for charge states $q_1$ and
$q_2$, respectively.

\subsubsection{Migration energies}

The migration barriers for single vacancies (\VO, \VBa, \VTi) were calculated
employing 40-atom supercells equivalent to $2\times 2\times 2$ unit
cells. Brillouin zone integrations were carried out using the same $k$-point
grids as for the formation energy calculations and the plane wave 
cutoff-energy was again 500\,eV. In order to obtain the saddle points,
we applied the climbing image nudged elastic band (CI-NEB) method
\cite{HenJohJon00, HenUbeJon00} which imposes a minimal number of
constraints on the transition path. Only jumps between nearest neighbor
sites were considered.

Unlike formation energies migration energies are obtained by
calculating differences between configurations which are structurally
and electronically very similar. As discussed in detail in
Ref.~\onlinecite{ErhAlb06a} the calculation of migration barriers is,
therefore, much less sensitive to the errors described in the
foregoing section. For this reason, comparably small supercells should
be sufficient and due to the similar electronic structure band gap and
potential corrections need not be applied.

\section{Results}

\subsection{Band structure}

In agreement with full potential-linearized augmented plane wave
\cite{SalHosSha04} and pseu\-do\-pot\-en\-tial plane wave
calculations, \cite{GhoGonMic99} our results predict the valence band
maximum (VBM) to be located at the R-point. On the other hand, in
previous tight-binding linear muffin-tin orbitals calculations
\cite{SahSinMoo00} within the atomic-sphere approximation the VBM was
located at the $\Gamma$-point. In all of these calculations the
difference between the highest occupied levels at R and $\Gamma$ was
calculated to be on the order of 0.1\,eV or less. The effective masses
of electrons and holes at the $\Gamma$-point, which can be used for
instance for a self-consistent determination of the Fermi level and
the charge carrier concentrations, \cite{ErhAlb07b} are given in
\tab{tab:methods_comparison}.

\subsection{Chemical potentials and stability diagram}
\label{sect:phasediagram}

In order to be able to derive defect formation energies from total
energy calculations, the thermodynamic reservoirs
need to be defined. This requires knowledge of the cohesive energies
of the constituents in their most stable conformation. Furthermore,
as the stability range of barium titanate is restricted by the
formation of competing phases such as TiO$_2$ or BaO, the formation
energies of these compounds have to be computed as well (see
\sect{sect:eform_calc}). The results of our calculations of the fully
relaxed structures are compiled in \tab{tab:phases} in comparison with
experimental data. The overall agreement with the reference data is
good. In particular, the calculated formation energies compare well with
experimental data.

The range within which the chemical potentials of Ba, Ti, and O can vary is
restricted by the condition
\begin{align}
   \Delta\mu_{\Ba} + \Delta\mu_{\Ti} + 3 \Delta\mu_{\O}
  &= \Delta H_f[\Ba\Ti\O_3],
  \label{eq:Hf_BTO}
\end{align}
which determines the outer triangle in the phase diagram shown in
\fig{fig:phasediagram}. Considering the compounds given in
\tab{tab:phases}, the chemical potentials are furthermore subject to
the following constraints
\begin{align*}
  \Delta\mu_{\Ba} + \Delta\mu_{\O}
  &\leq \Delta H_f[\Ba\O] \\
  \Delta\mu_{\Ba} + 2 \Delta\mu_{\O}
  &\leq \Delta H_f[\Ba\O_2] \\
  \Delta\mu_{\Ti} + 2 \Delta\mu_{\O}
  &\leq \Delta H_f[\Ti\O_2].
\end{align*}
They confine the stability range of BaTiO$_3$ to the gray shaded area
in \fig{fig:phasediagram}. Repeating this analysis with the
experimental data gives a phase diagram in good agreement with the
calculated one.

\subsection{Defect formation energies}
\label{sect:results_eform}

\begin{figure*}
  \centering
  \includegraphics[width=\linewidth]{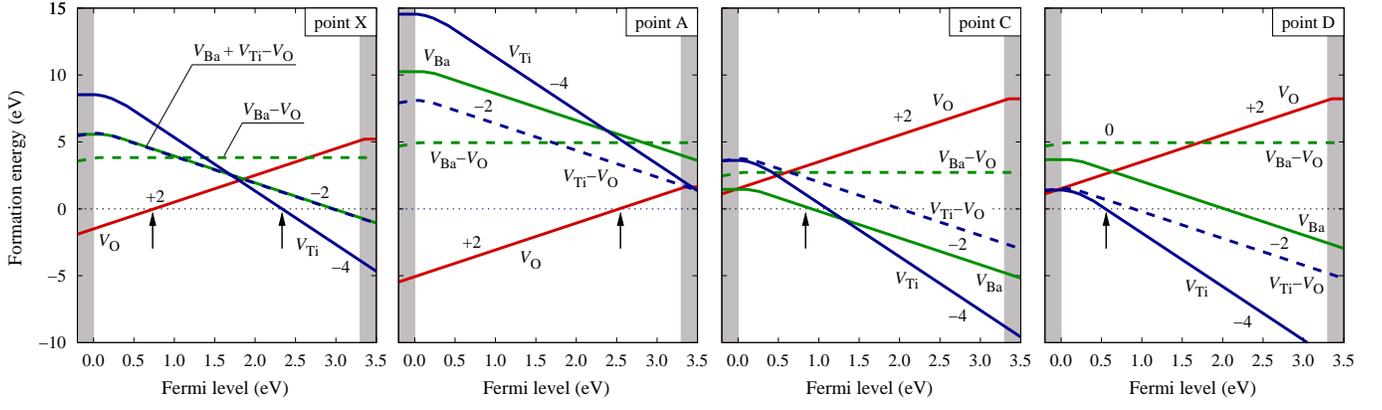}
  \caption{
    (Color online)
    Variation of defect formation energies with Fermi level for
    representative thermodynamic conditions indicated in
    \fig{fig:phasediagram}.
    The numbers indicate the charge states. Parallel lines correspond
    to identical charge states.
    The solid and dashed lines  correspond to mono and di-vacancies,
    respectively. The arrows indicate the position of the Fermi level
    pinning energy under different conditions.
  }
  \label{fig:eform}
\end{figure*}

The results of our calculations for the formation energies are
summarized in \tab{tab:eform} which shows the formation energies of
mono and di-vacancies for a Fermi level at the valence band maximum
($E_F=0\,\eV$ in equation \eq{eq:eform}) and for four representative
combinations of chemical potentials, which are indicated in
\fig{fig:phasediagram}. The variation with the Fermi level is shown
for the same four cases in \fig{fig:eform}.
Due to the large formation enthalpy of \BTO\ the formation energies
vary strongly between the extremal points of the phase diagram.
In the metal-rich limit (along A--B) oxygen vacancies
prevail. They have comparably small formation energies and therefore
should be abundant defects. In both cases the formation energies
becomes negative for some Fermi level which determines the so-called
pinning energy,
$\epsilon_{\text{pin}}$. This implies that under equilibrium
conditions the material cannot assume a Fermi level which is closer to
the valence band maximum than $\epsilon_{pin}$. \cite{ZhaWeiZun00} In
the oxygen-rich limit either barium (point C) or titanium vacancies
(point D) dominate. Fermi level pinning now occurs in the vicinity of
the conduction band, which implies that the Fermi level cannot be
pushed arbitrarily close to the conduction band minimum.

\begin{figure}
  \centering
  \includegraphics[width=0.75\linewidth]{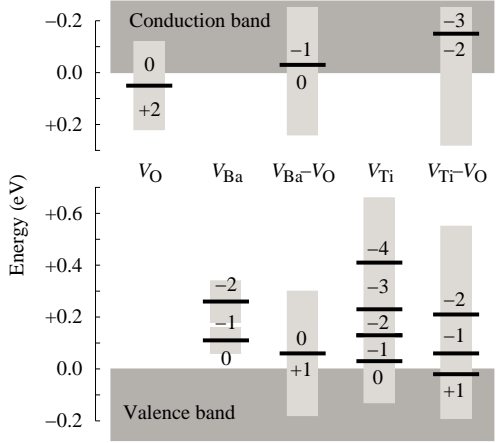}
  \caption{
    Transition levels for mono and di-vacancies in BaTiO$_3$. 
    Only the band edges are shown. The dashed transition levels are
    positioned inside the valence or the conduction bands (indicated
    by the light gray shaded areas) and are only included for
    illustration. The dark grey shaded areas indicate the sum of the
    extrapolation errors for each transition.
  }
  \label{fig:trans}
\end{figure}

The equilibrium defect transition levels can be deduced from the
formation energies using equation \eq{eq:trans}. They are presented in
an effective band scheme in \fig{fig:trans}. Since vacancies occur in
their nominal charge states ($V_{\Ti}^{''''}$, $V_{\Ba}^{''}$,
$V_{\O}^{\cdot\cdot}$) almost over the entire band gap, only the band
edges are shown.
%NEW: begin
With regard to the effect of the band gap correction term given by
equation \eq{eq:bandgap_correction}, it turns out that if no band gap
corrections are applied the donor transition levels ($q>0$) are near
the {\em calculated} conduction band edge, whereas they are near the
{\em experimental} conduction band edge if the corrections are
included. This consistency indicates that the application of the band
gap correction described in Ref.~\onlinecite{PerZhaLan05} is
reasonable in the present case.
%NEW: end

The location of the levels near the band edges is in agreement with
several models which have been developed to reproduce the experimental
data (see e.g., Refs.~\onlinecite{DanHar76, EroSmy78, ChaShaSmy81,
  ChaShaSmy82, ChaSmy84}). It is evident from the extrapolation errors
in \tab{tab:eform} that typically DFT calculations cannot predict
transition levels with an accuracy better than about 0.1\,eV. However,
even taking into account this limitation the present results show
clearly the intrinsic defect levels to be very close to the band edges
($\lesssim\!0.4\,\eV$). In particular, this applies for the titanium
vacancy for which, based on more simplistic calculations, the defect
level has been previously calculated to be located $-1.14\,\eV$ below
the conduction band maximum. \cite{LewCat86} A finding which is
confuted by the present results.

\begin{figure}
  \centering
  \includegraphics[width=0.8\linewidth]{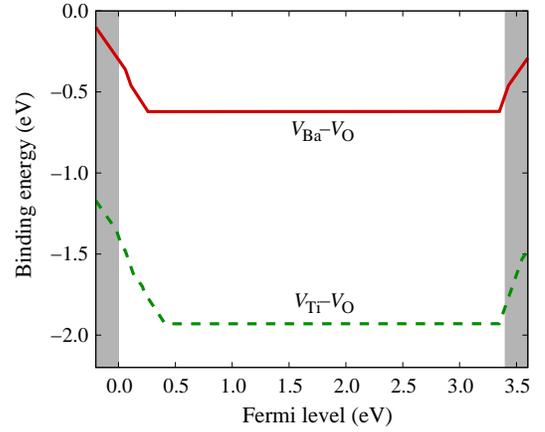}
  \caption{
    (Color online)
    Binding energies for \VBaVO\ and \VTiVO\ di-vacancies as a
    function of Fermi level. The kinks correspond to charge
    transition points of the isolated defects (compare
    \fig{fig:eform} and \fig{fig:trans}).
  }
  \label{fig:ebind}
\end{figure}

The binding energy of a di-vacancy is given by the difference between
its formation energy and the formation energies of the isolated
mono-vacancies. It is therefore independent of the chemical
potentials. Since charge transitions occur only within about
$0.4\,\eV$ of the band edges (see \fig{fig:trans}), the Fermi level
position has very little influence (see \fig{fig:ebind}). Over the widest
range of the band gap the binding energy of the \VTiVO-di-vacancy is
$E_b=-1.93\,\eV$ while the value for the \VBaVO-di-vacancy is
$E_b=-0.62\,\eV$. Only near the band edges the attraction are somewhat
reduced. Thus, the association of metal and oxygen vacancies is
energetically strongly favored, especially in the case of the
\VTiVO-di-vacancy.

Note that the formation energy of the \VBaVO-di-vacancy is constant
along lines which are parallel to A--D in the phase diagram
(\fig{fig:phasediagram}) corresponding to the reaction $\BTO +
\VBaVO \rightarrow \text{TiO}_2$. Equivalently, along lines
parallel to B--D the sum of the formation energies of oxygen vacancies
and \VTiVO-di-vacancies is constant, ($\BTO + \VTiVO + \VO
\rightarrow \text{BaO}$).

\subsection{Migration energies}
\label{sect:results_migration}

\begingroup
%\squeezetable
\begin{table}
  \caption{
    Calculated migration energies of mono-vacancies in
    units of eV. The temperature ranges above which the defects become
    mobile are given in the last column. The negative charge states of
    the titanium vacancy were not considered since already the neutral
    charge state displays a huge barrier and, following the trends for
    the barium and oxygen vacancies, the addition of electrons can
    only be expected to further increase this value.
  }
  \label{tab:migration}
  \centering

  \begin{tabular}{lddc}%x}{\linewidth}{Xddc}
    \hline\hline
    
    \tbhd{Defect}
    & \multicolumn{1}{c}{\tbhd{Charge}}
    & \multicolumn{1}{c}{\tbhd{Barrier}}
    & \tbhd{Onset of mobility}
    \\[3pt]

    \hline
    
    V$_{\Ba}$
    &  0 & 5.82 & $>2500\,\K$ \\
    & -1 & 5.96 & $>2600\,\K$ \\
    & -2 & 6.00 & $>2600\,\K$ \\[3pt]
    
    V$_{\Ti}$
    &  0 & 9.84 & $>4300\,\K$ \\[3pt]
    
    V$_{\O}$
    &  0 & 1.12 & 490--590\,K \\
    & +1 & 0.97 & 420--510\,K \\
    & +2 & 0.89 & 390--480\,K \\
    
    \hline\hline
    
  \end{tabular}

\end{table}
\endgroup

The calculated migration barriers are compiled in \tab{tab:migration}. The
smallest migration energies are obtained for oxygen vacancies and display a
weak charge state dependence. Experimentally the migration barrier for oxygen
vacancies has been determined to be $\Delta H_m=0.91$
(Ref.~\onlinecite{WarVanDim96}). The calculations compare well with this
value. In particular, the barrier for the doubly positive charge state, in
which the oxygen vacancy should occur for a Fermi level in the middle of the
band gap, is in very good agreement with this reference value. We
point out that the migration barriers for the different charge states of the
oxygen vacancy are very similar to the values for cubic lead titanate
calculated by Park within DFT. \cite{Par03}

Using the Einstein relation $6 D \tau = \left<r^2\right>$, one can
estimate the temperature above which a defect becomes mobile by
determining the temperature for which $6 D \tau$ exceeds
$\sqrt{\left<r^2\right>_{min}}$ (compare Ref.~\onlinecite{EhrJunSch91,
  ErhAlb06a}). The
pre-factor for the defect diffusivity can be approximated by the
lowest optical phonon mode at the $\Gamma$-point which gives
$\Gamma_0\approx 5\,\text{THz}$ (Ref.~\onlinecite{GhoGonMic97}). If
one assumes a typical isochronal annealing
time of $\tau=10\,\text{min}$ and a mean defect separation between
$\sqrt{\left<r^2\right>_{min}}=100\,\nm$ and $1000\,\nm$, one arrives 
at the values which are given in the last column of
\tab{tab:migration}. Obviously the only defects, which are fully mobile
at typical processing temperatures, are oxygen vacancies. In contrast, due to
their very large activation barriers, the migration of metal vacancies is
much lower even at temperatures close to the melting point ($\sim
1900\,\K$, Ref.~\onlinecite{crchandbook}).

\section{Discussion}

Experimentally, at low oxygen partial pressures (n-type region) and high
temperatures ($\sim\!1300-1500\,\K$) the dependence between the electrical
conductivity and the oxygen partial pressure is found to be $p_{\O_2}^{-1/6}$
which has been assigned to doubly charged oxygen vacancies. \cite{DanHar76,
  EroSmy78, ChaShaSmy81, ChaShaSmy82, ChaSmy84} In accordance the present
calculations predict oxygen vacancies to be by far the most important defect
under metal-rich conditions (low oxygen partial pressure) and to occur in
charge state $2+$ almost over the entire band gap. At somewhat lower
temperatures ($\lesssim 1300\,\K$) a transition to a $p_{\O_2}^{-1/4}$
dependence is observed. At least two different explanations have been
discussed in the literature. \cite{DanHar76, EroSmy78, ChaShaSmy81} Either (1)
the charge state of the oxygen vacancies changes from $2+$ to $1+$ or (2)
accidental acceptor dopants are present in the material. In order for
the first explanation to be valid, the $2+/1+$ transition level of the oxygen
vacancy should be located $1.3\,\eV$ below the conduction band minimum.
\cite{DanHar76} However, since the present calculations locate this
transition just 0.1\,eV below the conduction band minimum, they
provide support for accidental acceptor doping as the cause
for the change in slope. In fact, a more detailed investigation of the
relation between the conductivity and the oxygen partial pressure
shows that this mechanism can also explain the $p_{\O_2}^{1/4}$
dependence observed for higher oxygen partial pressures
($p_{\O_2}\gtrsim 10^{-2}\,\atm$, p-type region). \cite{ErhAlb07b}

The structural and energetic differences between the para-electric,
cubic phase and the ferroelectric phases play a crucial role in
determining the magnitude and the temperature dependence of
ferro-electricity. Since BaTiO$_3$ ceramics are
typically processed at temperatures above the cubic--tetragonal phase
transition ($T_c=393\,\K$) and since the onset of mobility
ranges given in \tab{tab:migration} exceed this temperature, the
point defect equilibria which are established during cooling
should correspond to the cubic phase. In particular, initially (prior
to ageing, see Refs.~\onlinecite{NeuArl87, ArlNeu88}) oxygen vacancies should
be rather randomly distributed over the symmetrically inequivalent oxygen
lattice sites, even if a clear energetic preference exists for a particular
lattice site (which is for instance the case for ``$c$-site'' vacancies in
tetragonal lead titanate \cite{ParCha98, ErhEicTra07}). In addition, due to
the large energy barriers for metal vacancy migration the distribution of
barium and titanium, which is established during growth, is expected
to be largely maintained if the material undergoes ferroelectric phase
transitions.

The binding energies for di-vacancies are negative and large which implies
a strong chemical driving force for their formation. Since the migration
barriers for metal vacancies are large, they are virtually immobile. In
contrast, oxygen vacancies are very mobile at temperatures
$\gtrsim\,500\,K$. Thus, formation of di-vacancies should occur readily at
typical growth temperatures by metal vacancies ``capturing'' diffusing oxygen
vacancies.

Defect complexes such as di-vacancies or impurity-vacancy associates carry a
dipole moment (see e.g., Ref.~\onlinecite{CocBur04} for a quantitative
calculation). On a cubic lattice different orientations of these defect
dipoles are energetically degenerate. In the presence of an electric field
\cite{NeuArl87, ArlNeu88} or for non-cubic lattices \cite{MesEicKlo05} this
degeneracy is, however, lifted (i.e., the energies for di-vacancy
pairs e.g., oriented along the $[001]$ and $[100]$ axes differs). As
argued above, depending on the barriers, the transition temperature
and the cooling rate defect dipoles might not have enough time to
achieve the orientation with the lowest energy. Since metal vacancies
are rather immobile, re-orientation of these dipoles is much more
likely to occur by oxygen vacancy jumps. The gradual reorientation of
defect dipoles is then determined by the barrier for oxygen vacancy
jumps in the first neighborhood of metal vacancies.

\section{Conclusions}

The thermodynamics and kinetics of vacancy defects in paraelectric
cubic barium titanate have been studied by means of density functional
theory calculations. First, formation, binding and migration energies
were derived properly accounting for the shortcomings of the DFT
method. The binding energies of metal-oxygen di-vacancies are found to
be negative implying that in equilibrium under metal-rich
(oxygen-rich) conditions all metal (oxygen) vacancies are bound in
di-vacancy clusters. While metal vacancies are practically immobile at
realistic conditions, oxygen vacancies can readily migrate at
typical growth temperatures. Di-vacancies can, therefore, form if
metal vacancies capture one oxygen vacancy.

It is furthermore confirmed that mono-vacancies occur in their
nominal (ionic) charge states ($V_{\Ti}^{''''}$, $V_{\Ba}^{''}$,
$V_{\O}^{\cdot\cdot}$) over the widest range of the band gap.
The dominant charge states for the di-vacancies are (\VTiVO)$^{''}$
and (\VBaVO)$^{\times}$. Intrinsic defect levels are confined to a
region within $\sim\!0.4\,\eV$ of the band edges. This is partially at
variance with earlier calculations based on a more simplistic model,
which suggested titanium vacancies to act as hole traps.
\cite{LewCat86}

The temperatures at which defects are immobilized are higher than or similar
to the cubic--tetragonal phase transition temperature. The principal
findings of the present study with respect to the thermodynamic
behavior of mono and di-vacancies are therefore also relevant for the
initial defect distribution in the ferroelectric phases.

\begin{acknowledgments}
This project was funded by the \textit{Sonderforschungsbereich 595}
``Fatigue in functional materials'' of the \textit{Deutsche
  Forschungsgemeinschaft}.
\end{acknowledgments}

\end{document}